\documentclass[letterpaper,aps,prl,twocolumn,showpacs]{revtex4}
\usepackage{graphicx,bm}
\usepackage{amsmath,amssymb}

\begin{document}

\title{Relativistic Ferromagnetic Magnon at the Zigzag Edge of Graphene}
\author{Jhih-Shih You, Wen-Min Huang and Hsiu-Hau Lin}
\affiliation{Department of Physics, National Tsing-Hua University, Hsinchu 30013, Taiwan}
\affiliation{Physics Division, National Center for Theoretical Sciences, Hsinchu 30013, Taiwan}
\date{\today}

\begin{abstract}
We study the spin-wave excitations near the zigzag edge of graphene. It is rather interesting that we obtain a single branch of relativistic ferromagnetic magnon due to the presence of the open boundary. Note that magnons in antiferomagnets appear in pairs, while the single brach magnon in ferromagnets does not have relativistic dispersion. Thus, the magnon near the zigzag edge of graphene is a hybrid of both, signaling its intrinsic property as a boundary excitation that must be embedded in a higher dimensional bulk system.
\end{abstract}

%82.39.Jn Charge (electron, proton) transfer in biological systems
%82.80.Gk Analytical methods involving vibrational spectroscopy
%87.16.Xa    Signal transduction and intracellular signaling
%87.19.lt    Sensory systems: visual, auditory, tactile, taste, and olfaction 

\maketitle

Graphene is the two-dimensional single-layer graphite composed of carbon atoms in honeycomb lattice. Recently, this novel material was successfully fabricated in laboratories\cite{Novoselov04,Novoselov05,Zhang05} and stimulates intense investigations on both experimental and theoretical sides. Its surprising band structure and high mobility seem promising for potential applications in many aspects\cite{Geim07}. For instance, there are proposals\cite{Rycerz07,Trauzettel07} that graphene nanoribbons can be the building block for future quantum computation. Since graphene is a low dimensional system, we expect the electron-electron interaction should play some role. In a recent paper, Son, Cohen and Louie\cite{Son06} spotted local magnetic moments near the zigzag edges of a graphene nanoribbon due to the Coulomb interaction. In fact, correlation-induced magnetic moment in graphene has be speculated for quite a while\cite{Fujita96,Nakada96,Wakabayashi99,Okada01,Hikihara03} but its low-energy excitations remain poorly understood at this point. 

\begin{figure}
\centering
\includegraphics[width=6cm]{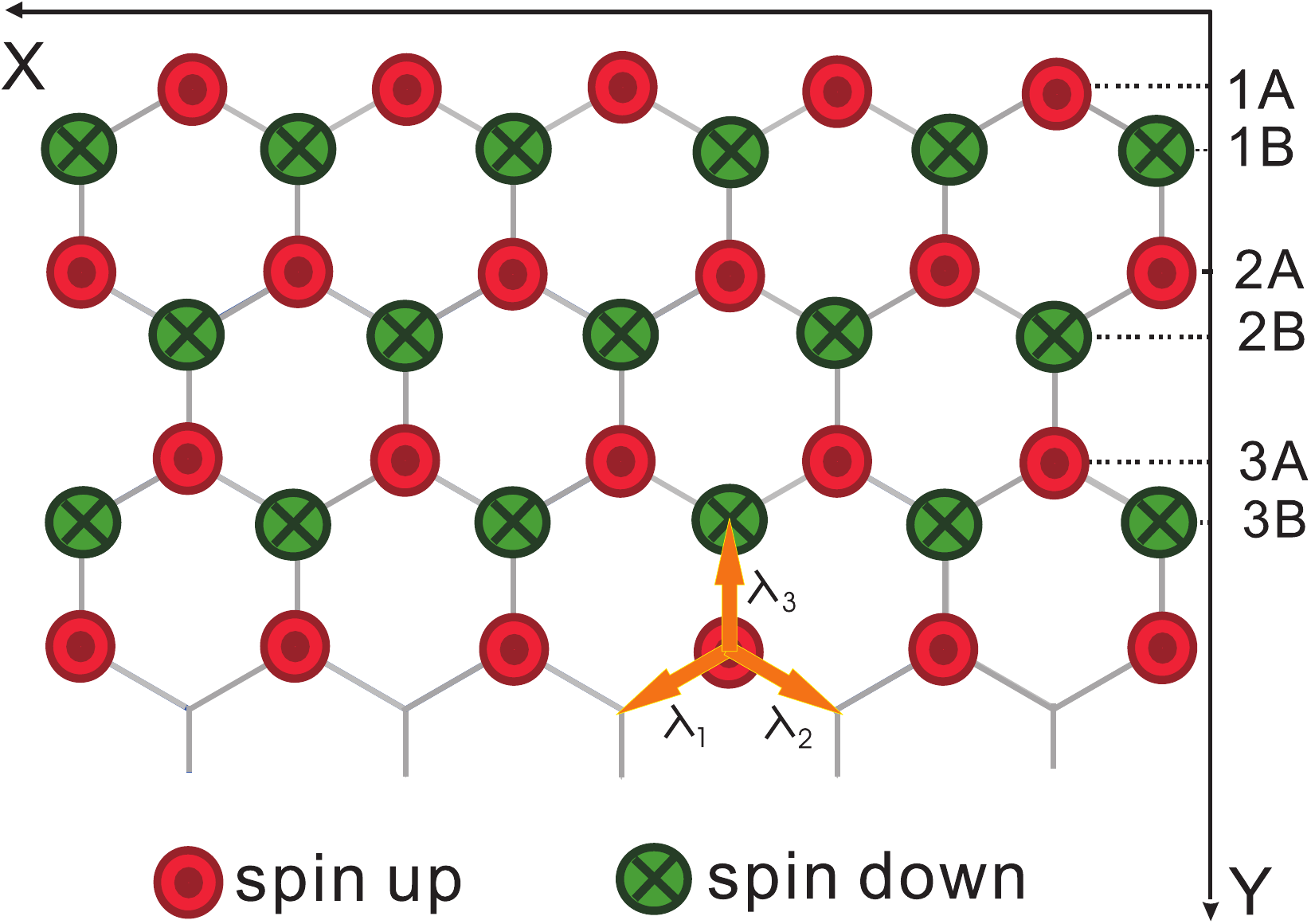}
\caption{Lattice structure for the semi-inifinite graphene with zigzag edge. The Neel state is shown by the alternative spins on different sublattices.}\label{neel}
\end{figure}

The novel magnetism shows that electronic correlations play a significant role at the edges of graphene. In condensed matter systems, the physical properties at the edge are often tied up with related bulk properties. For instance, the Andreev bound state\cite{Hu94} near the edge of a superconductor are related to the edge topology and also the pairing symmetry in the bulk. The edge states\cite{Wen90} of a quantum Hall liquid are described by the chiral Luttinger liquid that cannot be separated from the bulk due to gauge invariance. Therefore, we are inspired to investigate the low-energy spin excitations near the edge of a single-layer graphene here. For simplicity, we concentrate on a semi-infinite honeycomb lattice with zigzag edge as shown in Fig. \ref{neel}. Since the band structure of graphene is well approximated by the nearest-neighbor hopping $t$, the tight-binding model is sufficient. Furthermore, if we assume the electron-electron interaction can be capture by an effective on-site interaction $U$, it is natural to model the semi-infinite graphene by the Hubbard model at half filling $\langle n \rangle=1$. 

Unfortunately, it is known that the two dimensional Hubbard model cannot be solved exactly/reliably at current stage. Thus, approximations are inevitable. Since we are interested in the spin physics, we propose to concentrate on the effective Heisenberg exchange coupling in the system,
\begin{equation}
H=J \sum_{\langle \bf l, l' \rangle}
\textbf S({\bf l}) \cdot\textbf
S({\bf l'}),
\end{equation}
where ${\bf l}$ labels all lattice sites and $\langle \bf l, l' \rangle$ means nearest neighbor pairs. The coupling constant is $J = 4t^2 /U >0 $ in the strong interaction limit but is viewed as an independent parameter here. The honeycomb lattice of graphene can be separated into two sublattices $A$ and $B$. The outmost edge sites belongs to the sublattice $A$ in our convention as shown in Fig.~\ref{neel}. Starting from a Neel ordering state, it is convenient to rotate the spin on sublattice $B$ by $S_x\rightarrow
S_x$, $S_y\rightarrow -S_y$, and $S_z\rightarrow -S_z$ so that the semi-classical ground state become uniformly ferromagnetic. Following the standard procedures, we can represent the spin on the lattice sites by the Holstein-Primakov boson
\begin{eqnarray}\label{hp}
S^+_{L}({\bf r}) &=& \sqrt{2S-b_{L}^\dag({\bf r}) b_{L}({\bf r})}\: 
b_{L}({\bf r}),
\nonumber\\
S_{L}^-({\bf r}) &=& b_{L}^\dag ({\bf r})\: 
\sqrt{2S-b_{L}^\dag({\bf r}) b_{L}({\bf r})},
\nonumber\\
S_{L}^z({\bf r}) &= & S - b_{L}^\dag({\bf r}) b_{L}({\bf r}),
\end{eqnarray}
where $b_{L}({\bf r}), b_{L}^\dag({\bf r})$ are annihilation/creation operators for bosons. To make the sublattice dependence explicit, we have separate the lattice positions into sublattice index $L = A, B$ and ${\bf r}$ for the triangular lattice. Expanding the Heisenberg interaction to the quadratic order in the bosonic operator, one can obtain the familiar Hamiltonian for spin-wave excitations,
\begin{eqnarray}
H&=&\nonumber JS\sum_{{\bf r},\bm{\lambda}} \bigg[
b_{A}^{\dagger}(\textbf{r})b_{A}(\textbf{r})+
b_{B}^{\dagger}(\textbf{r}+\bm{\lambda})b_{B}(\textbf{r}+\bm{\lambda})\\
&&+ b_{A}^{\dagger}(\textbf{r})b_{B}^{\dagger}(\textbf{r}+\bm{\lambda})+
b_{B}(\textbf{r}+\bm{\lambda})b_{A}(\textbf r) \bigg].
\end{eqnarray}
The summation over $\bm{\lambda}$ includes all three nearest neighbors 
$\bm{\lambda}_i = a(\pm 1/2,1/2\sqrt{3}), a (0,-1/\sqrt{3})$ as shown in Fig.~\ref{neel}. The presence of open boundary ruins the translational invariance in $y$ direction and the above Hamiltonian cannot be diagonalized by the usual Fourier transformation.

However, since the translational invariance in the $x$ direction is still valid, we can perform the partial Fourier transformation
\begin{eqnarray}
b_{L}(x,y)=\frac{1}{\sqrt N_x}\sum_{k_x}e^{ik_{y}y} \psi_{L}(k_x,y),
\end{eqnarray}
to simplify the Hamiltonian. After some algebra, the spin-wave Hamiltonian can be casted into the form,
\begin{eqnarray}
H  &=& JS\sum_{k_x} \Psi^\dagger \begin{bmatrix}
  h_{11}& h_{12}  \\
  h_{21} & h_{22}
\end{bmatrix} \Psi,
\end{eqnarray}
where we introduce the two-component spinor
$\Psi^\dag = [\psi^\dag_A(k_x,y), \psi_B(-k_x,y)]$.
The matrix elements in the reduced $2 \times 2$ matrix are semi-infinite matrices: $h_{11}=2 +D^\dag D$, $h_{22}=3 $ and $h_{21} = h^{\dagger}_{12} = J_1 + J_2 D$, where $J_1=2\cos{\frac{k_x a}{2}}$ comes from the tilted bonds and $J_2=1$ from the vertical bonds. For anisotropic interaction, $J_2$ would derivate from one. The semi-infinite displacement matrix $\textbf{D}$ is
\begin{eqnarray*}
D=\begin{bmatrix}
0 & 1 & 0 & 0  & \cdots\\
0 & 0 & 1 & 0  & \cdots \\
0 & 0 & 0 & 1  & \cdots \\
0 & 0 & 0 & 0   & \cdots\\
\cdots & \cdots & \cdots & \cdots & \cdots
\end{bmatrix}.
\end{eqnarray*}

Now the problem in converted into solving the eigenstates for a one dimensional semi-infinite system. For notation clarity, we suppress the $k_x$ momentum dependence in the following. Writing down the matrix elements of the Hamiltonian explicitly, it leads to the coupled Harper equations,
\begin{eqnarray}
3\psi_{A}(n)+J_1\psi_{B}(n)+J_2\psi_{B}(n-1)&=&\varepsilon\psi_{A}(n)
\nonumber\\
-3\psi_{B}(n)-J_1\psi_{A}(n)-J_2\psi_{A}(n+1)&=&\varepsilon\psi_{B}(n),
\end{eqnarray}
where $n=1,2,3,...$ denote the lattice coordinates in the $y$ direction. However, one needs to pay special attentions to the boundary. For zigzag edge, there is one missing bond from the outmost $A$ sites to the opposite sublattice. The semi-infnite nature of the Hamiltonian gives rise to the constraint,
\begin{eqnarray}\label{b1}
\psi_{A}(1)+J_2\psi_{B}(0)=0.
\end{eqnarray}

Now we are ready to find out the eigenstates by the generalized Bloch theorem\cite{Lin05}. Note that eigenvalue of a unitary displacement operator can be written as $z = e^{ik}$, where $k$ is the momentum. However, the presence of open boundary changes to this nice connection to the momentum. It is easy to check that the displacement operator satisfies $DD^\dag=1$ but not in the inverse order $D^\dag D \neq 1$. Thus, it is no longer unitary. Thus, the eigenstate should be written in the more general form,
\begin{eqnarray}
\Psi =\begin{bmatrix}
   \psi_{A}(n) \\
   \psi_{B}(n)
   \end{bmatrix}
   =\begin{bmatrix}
  c_{A} \\
 c_{B}
   \end{bmatrix}z^n.
\end{eqnarray}
Upon substitution into the coupled Harper equations, the eigenvalue $z$ of the displacement operator satisfy the simple algebraic constraint,
\begin{eqnarray}\label{ch1}
J_{1}J_{2}\left(z+\frac{1}{z}\right)+(J_{1}^2+J_{2}^2+{\varepsilon}^2-9)=0.
\end{eqnarray}
Usually, for given momentum $k$, we look for energy $\varepsilon$ of the system. Here it is more convenient to look for the complex $z$ for a given energy $\varepsilon$. It is obvious that the solutions $z, 1/z$ appear in pairs. Thus, the solutions can be classified into two types: (1) plane-wave solutions with $z = e^{\pm ik}$ (2) evanescent modes with real $z = \alpha, 1/\alpha$. Since we require the wave function to be finite at infinity, only one of the evanescent solutions survives while, not surprisingly, both plane-wave solutions are allowed.

\begin{figure}
\centering
\includegraphics[width=7cm]{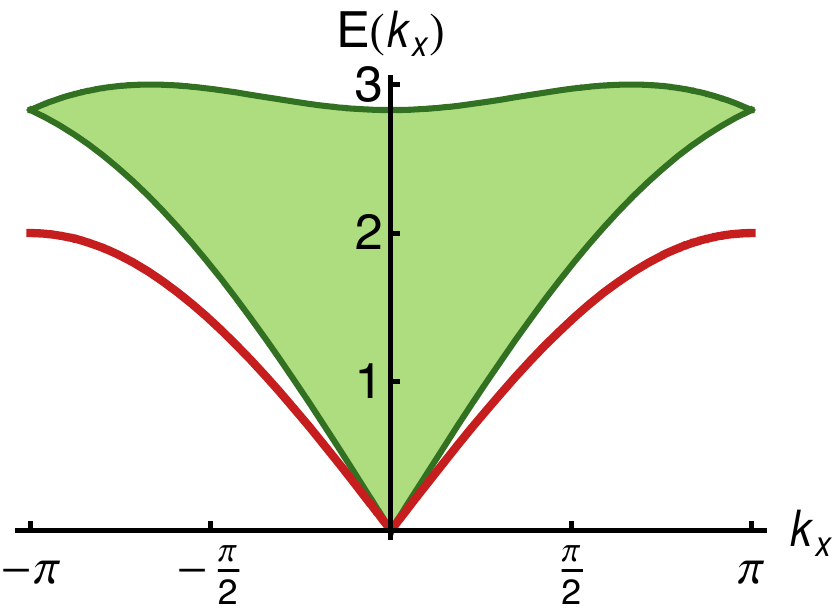}
\caption{Dispersion for bulk and edge magnons with $JS=1$ and $a=1$ simplicity. The shaded area represents the bulk magnons with double degeneracy $\Delta S^z = \pm 1$. The thick blue line indicates the dispersion for the edge magnon with only one brach $\Delta S^z = -1$.
}\label{fig:band}
\end{figure}

The plane-wave solutions correspond to the bulk magnons. The presence of the open boundary does cause a rather complicated mixing between counter propagating modes
$\pm k_y$. However, the energy spectrum remains the same as shown in Fig.~\ref{fig:band}. Therefore, we would only concentrate on the more interesting evanescent modes where $z$ is real. Solving the coupled Harper equations together with the boundary condition, there exist only one solution $\varepsilon = \sqrt{4-J_1^2}$. Therefore, the dispersion for the single-branch edge magnon is
\begin{eqnarray}
E_{e}(k_x) = 2 JS |\sin (k_x a/2)|,
\end{eqnarray}
which becomes linear (relativistic) in the small momentum regime. Note that, as shown in Fig.~\ref{fig:band}, the velocity of the edge magnon is slightly smaller than that for the bulk magnon and thus is protected by a gap. In fact, this protection arises from the sharp distinction between $|z|=1$ and $|z|<1$ solutions. Furthermore, the wave function of the edge magnon is
\begin{eqnarray}
\Psi = \begin{bmatrix}
1/z \\
-1
\end{bmatrix} z^n, \qquad \mbox{with\hspace{4mm}}
z = \frac{1-|\sin(k_x a)|}{\cos(k_x a/2)}.
\end{eqnarray}
Note that $0 \leq z < 1$ as required for the evanescent modes except at $k_x=0$ where the edge magnon merges into the bulk with $z=1$.

\begin{figure}
\centering
\includegraphics[width=8cm]{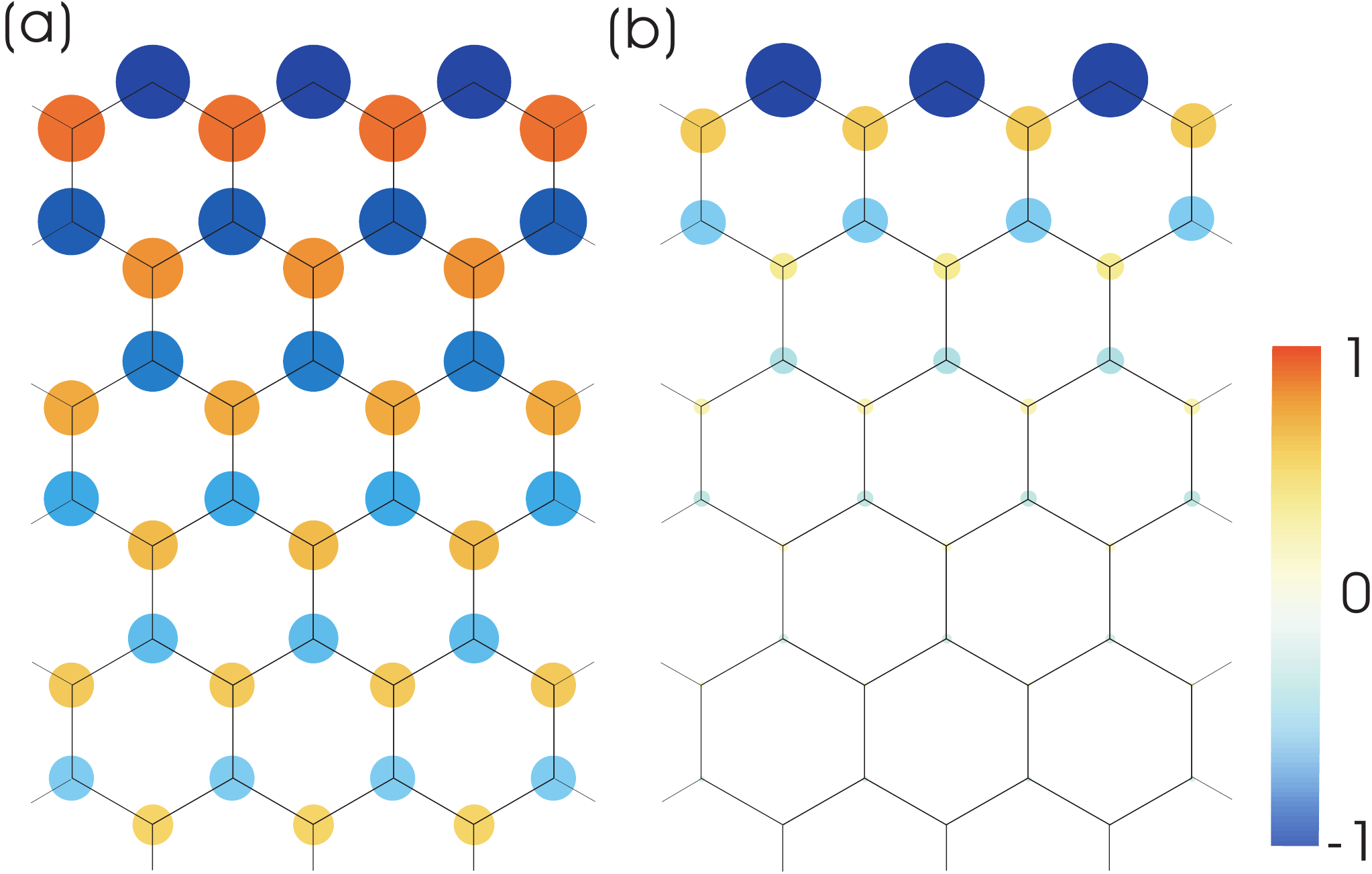}
\caption{Spin density profile for (a) $k_x = 0.1$ and (b) $k_x = 0.5$ edge magnons. For clarity, the magnitudes on the edge are held constant. The magnitudes of the spin density (after subtracting the background from the ground state)  are represented by the radius and also the color gradients.}\label{fig:spin}
\end{figure}
 
The spin density profile of the edge magnon can be obtained by subtracting the background form the ground state, $\Delta S^z( \textbf r)\equiv\langle1|S^z(\textbf r)|1\rangle-\langle G|S^z(\textbf r)|G\rangle,$ where $|1\rangle$ is one magnon state and $|G\rangle$ is the ground state. Since the spin density is uniform in the $x$ direction, a factor $1/N_x$ appears where $N_x$ is the total transverse length. The spin density can be expressed in terms of the normalized wave functions,
\begin{eqnarray}
\Delta S^z(\textbf r)=\left\{\begin{array}{ll}
                 -\frac{1}{N_x}|\psi_A(n)|^2, &  \mbox{ $\textbf r\in A$,} \\
                +\frac{1}{N_x}| \psi_B(n)|^2, & \mbox{ $\textbf r\in
                B$}.
                \end{array} \right.
\end{eqnarray}
Summing over all lattice sites, it is easy to check the edge magnon we found carries the total spin $\Delta S^z = \sum_{\bf r} \Delta S^z({\bf r}) = -1$, as for those found in conventional ferromagnets. The spin density profiles of the edge magnon with two different momenta $k_x=0.1$ and $k_x=0.5$ are shown in Fig.~\ref{fig:spin}. For momentum fairly close to $k_x=0$, the spin density extends into the bulk with long decay length. However, for $k_x=0.5$, though not very far away from the origin, the spin profile is already quite localized to the zigzag edge.

So, what's the significance of the above simple calculations? For an ordinary ferromagnet, there is one branch of non-relativistic magnon. On the other hand, antiferromagnetic ground state supports two branches of relativistic magnons. Here, near the zigzag edge of graphene, we find a hybrid between the two cases -- a single branch of the relativistic ferromagnetic magnon. This implies that the spin-wave excitations near the edge of the graphene cannot be described as the ordinary one dimensional ferromagnet and shall be viewed as an intrinsic boundary field theory. The missing branch can be understood by the following symmetry argument. Although an antiferromagnetic order breaks both time-reversal and sublattice exchange symmetries separately, the ground state remains invariant under both transformations simultaneously. Thus, starting from the $\Delta S^z = -1$ branch, one can reverse the spin orientations by time-reversal transformation followed by sublattice exchange to construct the other $\Delta S^z =1$ branch. This explains why antiferromagnetic magnons on bipartite lattices always appear in pairs with opposite spins. However, the sublattice symmetry is broken in the presence of the zigzag edge, this makes the single-branch magnon possible. In fact, we check with the armchair edge which doesn't destroy the sublattice symmetry and found no evidence for the novel edge magnon discussed here.

Finally, we address the validity of the approximations in the linearized spin-wave theory. First of all, the charge excitations are all left out. Since there is no charge gap in graphene, it is expected that the neutral particle-hole excitations with $\Delta S^z=-1$ will mix up with the edge magnon. However, since the Fermi surface shrinks to two Dirac points, the phase space is largely reduced. Furthermore, since the Fermi velocity $v_{F}$ is expected to be faster than the magnon velocity $v_m$, the hybridization is further suppressed by a finite gap $\Delta \sim (v_{F}-v_{m}) k_x$ for any finite momentum. This is similar to the protection between bulk and edge magnons we discussed previously. Another approximation is the existence of the Neel order. This is the most serious approximation since it is likely that the antiferromagnetic order may not exist at all in graphene. Though the quadratic fluctuations from the magnons do not destroy the Neel order and the spin-wave theory remain self-consistent, it remains an open question whether our findings here are true beyond spin-wave theory. Given the novelty of the edge magnetism in graphene, alternative theoretical investigations are in oder. And, as always, more experimental data will help to clear up our understanding of spin-wave excitations living on the edge.

We acknowledge supports from the National Science Council in Taiwan through grants NSC-96-2112-M-007-004 and NSC-97-2112-M-007-022-MY3. Financial supports and friendly environment provided by the National Center for Theoretical Sciences in Taiwan are also greatly appreciated.


\begin{thebibliography}{99}

% Fabrication of graphene
\bibitem{Novoselov04}
K. S. Novoselov, A. K. Geim, S. V. Morozov, D. Jiang, Y. Zhang, S. V. Dubonos, V. Grigorieva and A. A. Firsov,
Science {\bf 306}, 666 (2004).

\bibitem{Novoselov05}
K. S. Novoselov, A. K. Geim, S. V. Morozov, D. Jiang, M. I. Katsnelson, I. V. Grigorieva, S. V. Dubonos and A. A. Firsov, 
Nature \textbf{438}, 197 (2005).

\bibitem{Zhang05}
Y. Zhang, Y.-W. Tan, H. L. Stormer and P. Kim,
Nature \textbf{438}, 201 (2005).

\bibitem{Geim07}
A. K. Geim and K. S. Novosselov, 
Nature Mat. \textbf{6}, 183 (2007) and references therein.

\bibitem{Rycerz07}
A. Rycerz, J. Tworzydlo and C. W. J. Beenakker, 
Nature Phys. \textbf{3}, 172 (2007).

\bibitem{Trauzettel07}
B. Trauzettel, D. S. Bulaev, D. Loss and G. Burkard, 
Nature Phys. \textbf{3}, 192 (2007).

\bibitem{Son06}
Y.-W. Son, M. L. Cohen and S. G. Louie, 
Nature \textbf{444}, 347 (2006).

% Localized state at zigzag edge
\bibitem{Fujita96}
M. Fujita, K. Wakabayashi, K. Nakada and K. Kusakabe,
J. Phys. Soc. Jpn. \textbf{65}, 1920 (1996).

\bibitem{Nakada96}
M. Fujita, K. Nakada, G. Dresselhaus and M. S. Dresselhaus,
Phys. Rev. B \textbf{54}, 17954 (1996).

\bibitem{Wakabayashi99}
K. Wakabayashi, M. Fujita, H. Ajiki and M. Sigrist,
Phys. Rev. B \textbf{59}, 8271 (1999).

\bibitem{Okada01}
S. Okada and A. Oshiyama,
Phys. Rev. Lett. \textbf{87}, 146803 (2001).

\bibitem{Hikihara03}
T. Hikihara, X. Hu, H.-H. Lin and C.-Y. Mou,
Phys. Rev. B \textbf{68}, 035432 (2003).

\bibitem{Hu94}
C. R. Hu,
Phys. Rev. Lett. {\bf 72}, 1526 (1994).

\bibitem{Wen90}
X. G. Wen,
Phys. Rev. Lett. {\bf 64}, 2206 (1990).

\bibitem{Kobayashi05} 
Y. Kobayashi, K. Fukui, T. Enoki, K. Kusakabe and Y. Kaburagi,
Phys. Rev. B {\bf 71}, 193406 (2005). 

\bibitem{Niimi06}
Y. Niimi, T. Matsui, H. Kambara, K. Tagami, M. Tsukada and H. Fukuyama,
Phys. Rev. B {\bf 73}, 085421 (2006).


\bibitem{Son06a}
Y.-W. Son, M. L. Cohen, and S. G. Louie,
Nature {\bf 444}, 347 (2006).

\bibitem{Son06b}
Y.-W. Son, M. L. Cohen and S. G. Louie,
Phys. Rev. Lett. {\bf 97}, 216803 (2006).

\bibitem{Brey07}
L. Brey, H. A. Fertig, and S. Das Sarma,
Phys. Rev. Lett. {\bf 99}, 116802 (2007).

\bibitem{Lin05}
T. Pereg-Barnea and H.-H. Lin,
Europhys. Lett. {\bf 69}, 791 (2005).

\end{thebibliography}
\end{document}